\documentclass[cits]{PoS}
\pdfoutput=1
\usepackage{booktabs,amsmath,amssymb}
\usepackage[utf8]{inputenc}

\newcommand{\ps}{\text{ps}}
\newcommand{\vt}{\text{vt}}

\newcommand{\res}{\text{res}}

\makeatletter
     {\bgroup\raggedright\small\section*{\refname
        \@mkboth{\MakeUppercase\refname}{\MakeUppercase\refname}}%
      \list{\name{bib\@arabic\c@enumiv}% HOPE!
	    \@biblabel{\@arabic\c@enumiv}}%
           {\settowidth\labelwidth{\@biblabel{#1}}%
            \setlength\itemsep{0pt}
            \setlength\parskip{0pt}
            \setlength\parsep{0pt}
            \setlength\partopsep{0pt}
            \leftmargin\labelwidth
            \advance\leftmargin\labelsep
            \@openbib@code
            \usecounter{enumiv}%
            \let\p@enumiv\@empty
            \renewcommand\theenumiv{\@arabic\c@enumiv}}%
      \sloppy\clubpenalty4000\widowpenalty4000%
      \sfcode`\.\@m}
     {\def\@noitemerr
       {\@latex@warning{Empty `thebibliography' environment}}%
      \endlist\egroup}
\makeatother

\title{Constructing a composite Higgs model \newline with built-in large separation of scales}

\ShortTitle{Constructing a composite Higgs model with built-in large separation of scales}

\author{\speaker{Oliver Witzel}\\
        Department of Physics, University of Colorado, Boulder, CO 80309, USA\\
        E-mail: \email{oliver.witzel@colorado.edu}}

\author{Anna Hasenfratz\\
        Department of Physics, University of Colorado, Boulder, CO 80309, USA\\
        E-mail: \email{anna.hasenfratz@colorado.edu}}  

\author{(Lattice Strong Dynamics collaboration)}
  
\abstract{Experimentally the existence  of a light 125 GeV Higgs boson is well established but so far no other heavier resonances have been observed. Viable models to describe the Higgs boson 
as composite particle require hence to exhibit a large separation of scales. This occurs naturally in systems located near a conformal fixed point irrespective whether the system lies outside or inside the conformal window. 
We demonstrate the latter case by investigating a mass-split model with four light and
six heavy flavors. By construction mass-split models exhibit a large separation of
scales and feature in addition a highly constrained hadron spectrum. We present results
based on the low-lying connected meson spectrum. Although the light sector is chirally broken, we show that it exhibits hyperscaling which is typical for conformal systems.} 
  
\FullConference{37th International Symposium on Lattice Field Theory - Lattice2019\\
		16-22 June 2019\\
		Wuhan, China}

\begin{document}

\section{Introduction}
The motivation of composite Higgs models is to provide a description of physics beyond the Standard Model (SM) with the aim to explain the origin of the Higgs boson and how electro-weak symmetry breaking arises. In such models the Higgs boson is a bound state of a new strongly interacting sector which obtains its observed quantum numbers by embedding the SM in the new strongly interacting sector. Experimental constrains rule out that the new strongly coupled sector is QCD-like. They also require that it exhibits a large separation of scales to explain why a light, 125 GeV Higgs boson \cite{Aad:2015zhl} has been observed by Atlas and CMS  \cite{Aad:2012tfa,Chatrchyan:2012ufa} but no other, heavier resonances. Commonly, two scenarios for a composite Higgs boson are considered:
\begin{itemize}
\item[A)] The Higgs boson is a scalar excitation of the strong sector. It could e.g.~be a dilaton arising from the breaking of the conformal symmetry or, generally, a $0^{++}$ bound state with properties similar to that of a dilaton. To be a viable candidate for the Higgs boson, the pseudoscalar of the strong interactions must be massless and the $0^{++}$, the lightest massive particle, must be much lighter than e.g.~the vector resonance. This excludes a QCD-like scenario because the $0^{++}$ ($f_0(500)$ or $\sigma$) with a mass of about 500 MeV is too heavy compared to the vector ($\rho$) with mass of about 770 MeV.
  \item[B)] Alternatively the Higgs boson is a pseudo-Nambu-Goldstone boson (pNGB) of the strong sector which acquires its mass solely from the interactions with the SM. A pNGB Higgs boson would arise similarly to a pion in massless QCD. A large scale separation is still required to generate the SM fermion masses either through partial compositness or 4-fermion interaction \cite{Vecchi:2015fma,Ferretti:2016upr}.
\end{itemize}
To create a system with the desired dynamics resulting in a large separation of scales, it is essential to have a slowly evolving (walking) coupling. In addition large anomalous dimensions are required to satisfy other phenomenological constraints. Both walking and large anomalous dimensions could occur near the opening of the conformal window. Identifying the number of flavors marking the onset of the conformal window is a challenging task and depends on the gauge group as well as the representation of the fermions. Although SU(3) gauge theories with fermions in the fundamental representation are particularly well studied (see e.g.~\cite{Deuzeman:2008sc,Hasenfratz:2014rna,Hasenfratz:2016dou,Ishikawa:2015iwa,Chiu:2016uui,Chiu:2017kza,Chiu:2018edw,Fodor:2017gtj,Fodor:2018tdg,Hasenfratz:2017qyr,Hasenfratz:2019dpr,Witzel:2019jbe,Hasenfratz:2019hpg,Hasenfratz:2019puu}), the onset of the conformal window is still controversial.

Here we report on a different approach that is appropriate to describe both dilaton-like and pNGB Higgs scenarios. Our setup of a mass-split system exhibits by construction a large separation of scales without requiring to sit just below the silt of the conformal window. Assuming the system with degenerate and massless $N_f$ flavors is conformal, we split the $N_f$ flavors into $N_\ell$ ``light''of mass $m_\ell$  and $N_h$ ``heavy'' flavors of mass $m_h$. The light flavors are  massless, whereas the heavy flavors retain their mass. The heavy flavor mass, however, has to be small enough such that in the ultraviolet (UV) energy range the system feels the attraction of the conformal fixed point which corresponds  to the theory with degenerate $N_f$ flavors. Mass-split systems build on a conformal fixed point have two remarkable properties \cite{Brower:2014dfa,Brower:2014ita,Brower:2015owo,Hasenfratz:2016gut}
\begin{itemize}
\item \textbf{Highly constrained particle spectrum:} Due to the presence of a conformal fixed point in the UV, the system exhibits conformal hyperscaling.  As a result physical observables depend only on the ratio of the light and heavy flavor masses, reducing the number of independent parameters of the system. Dimensionless ratios of hadron masses or hadron masses over amplitudes are thus independent of the mass of the heavy flavors. Since we are interested in taking the chiral limit of the light sector (i.e.~$m_\ell \to 0$), and the gauge coupling is an irrelevant parameter in the vicinity of the conformal IRFP, the mass of the heavy flavors is the only free parameter of mass-split systems.  Thus $m_h$ takes over the role of the gauge coupling of QCD-like systems and, typically,  the pseudoscalar decay constant of the light sector $F_\ps^{\ell\ell}$, is used to set the scale.  
\item \textbf{Tunable range of walking dynamics:} Although the particle spectrum shows hyperscaling and the mass of the heavy flavors only enters through the ratio $m_\ell/m_h$,  tuning $m_h \to 0$ increases the range of a walking coupling.
\end{itemize}
In addition, a further consequence of near-conformal dynamics seems to be that the iso-singlet scalar is the lightest, massive particle in the chiral limit. This observation is based on numerical data determining the low-lying particle spectrum \cite{Brower:2015owo,Weinberg:2014ega,Aoki:2013xza,Aoki:2016wnc,Appelquist:2016viq,Appelquist:2018yqe}.

In the following  we focus on the details of our mass-split system with SU(3) gauge group, four light and six heavy flavors \cite{Witzel:2018gxm}.\footnote{Although the nature of SU(3) with ten fundamental flavors has not been conclusively resolved \cite{Chiu:2016uui,Chiu:2017kza,Chiu:2018edw,Fodor:2017gtj,Fodor:2018tdg,Hasenfratz:2017qyr}, we consider $N_f=10$ to have properties resembling quite closely a conformal theory and provide numerical support for this later.} Choosing a system with four light flavors is e.g.~motivated by Two-Higgs-Doublet model by Ma and Cacciapaglia \cite{Ma:2015gra}, whereas a scenario with ten flavors is e.g.~outlined by Marzocca \cite{Marzocca:2018wcf} or part of the considerations by Vecchi \cite{Vecchi:2015fma}.  For this study we generated a set of dynamical gauge field configurations with three times stout smeared M\"obius domain wall fermions (MDWF) and Symanzik gauge action \cite{Kaneko:2013jla,Noaki:2015xpx}. Our analysis includes simulations on $32^3\times 64$ and $24^3\times 64$ volumes. We simulate MDWF using a domain wall height $M_5=1$ and set the extent of the fifth dimension to $L_s=16$. Choosing three values for the mass of the heavy flavor, $a m_h=0.200$, 0.175, and 0.150, we select in each case three different light flavor masses:
\begin{align*}
  am_h=0.200: &\qquad am_\ell=0.015,\, 0.020, \text{ and } 0.030\\
  am_h=0.175: &\qquad am_\ell=0.018,\, 0.026, \text{ and } 0.035\\
  am_h=0.150: &\qquad am_\ell=0.015,\ 0.023, \text{ and } 0.033  
\end{align*}
In addition we show for illustrational purpose two results obtained with $am_h=0.200$ on $16^3\times 32$ volumes using $am_\ell=0.040$ and 0.050. While these heavier data points confirm in general our observations, the small volumes show  finite volume corrections and likely increased systematic effects. In the future we intend to replace these data point by simulations on larger $24^3\times 64$ lattices.

\begin{figure}[tb]
  \centering
  \includegraphics[height=0.165\textheight]{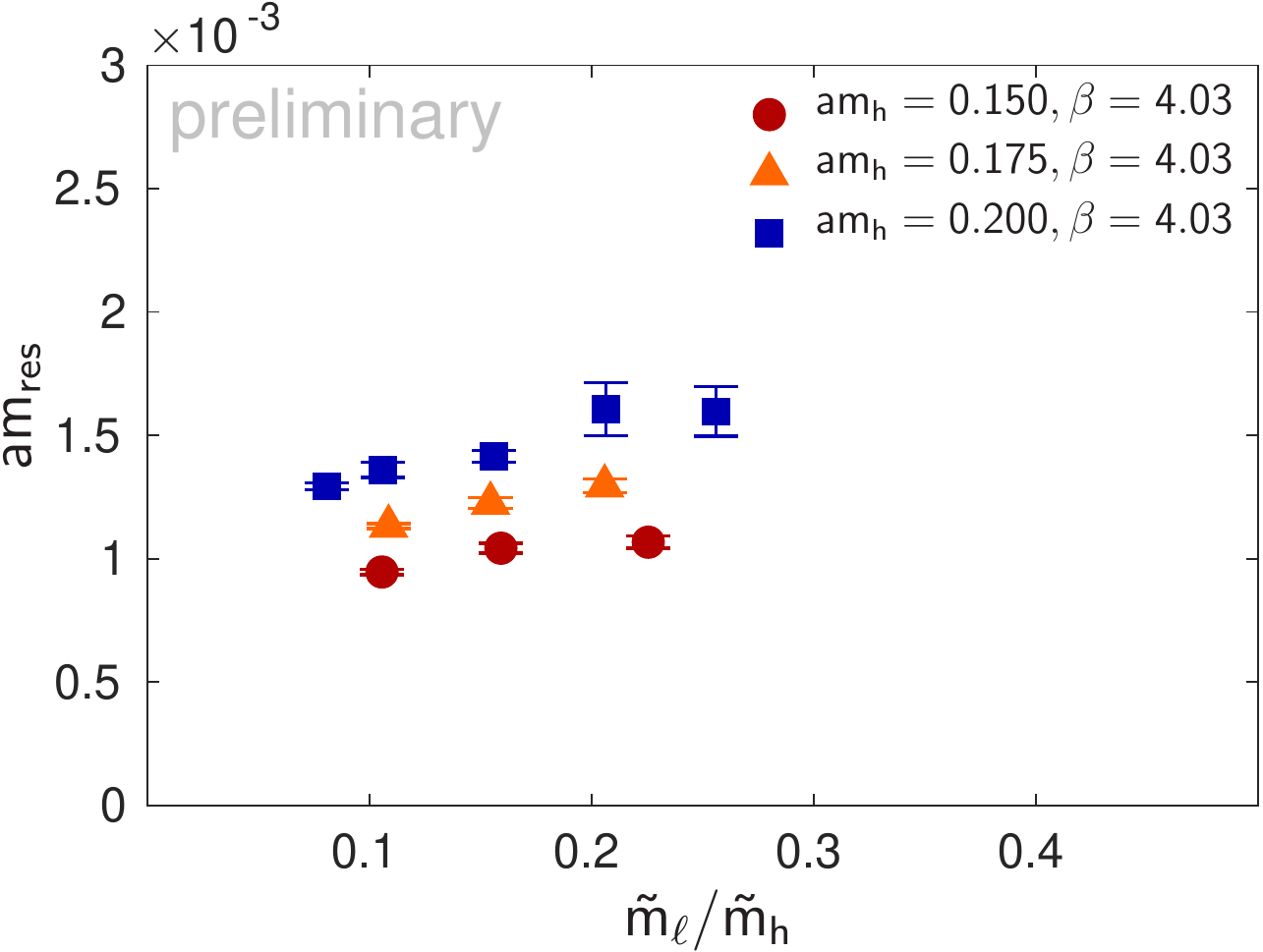}
  \caption{Residual chiral symmetry breaking expressed as a residual mass term $am_\res$ numerically determined using the midpoint-pseudoscalar and pseudoscalar-pseudoscalar correlator. Only statistical uncertainties are shown.}
  \label{fig.mres}
\end{figure}

We note that for all ensembles we find an average plaquette, normalized to one, very close to 0.57. This means that all our configurations are relatively smooth despite simulating with very strong bare coupling of $\beta=4.03$. In addition we verify the chiral properties by calculating the residual mass $am_\res$ which parameterizes the residual chiral symmetry breaking present in MDWF due to the finite extent of the fifth dimension.  Using the midpoint-pseudoscalar and the pseudoscalar-pseudoscalar correlator we determine $am_\res$ for all ensembles and show the results in Fig.~\ref{fig.mres}. The value of $am_\res$ decreases slightly for decreasing values of $am_h$ but is for all simulations sufficiently small and of order $10^{-3}$.

In the following section we present results of our analysis of the connected meson spectrum before we close with a brief outlook.

\section{Connected meson spectrum}
We determine the connected spectrum using all presently available gauge field configurations separated by 20 molecular dynamics time units (MDTU) i.e.~we analyze 130 to 270 gauge field configuration per ensemble and further decorrelate the measurements by performing a random 4-vector shift of the gauge field before placing the sources. On each configuration we place a Z2 wall source \cite{Boyle:2008rh} every eight time slices and start the subsequent data analysis by first averaging correlators on the same configuration. We account for residual autocorrelation by estimating the statistical uncertainties using the $\Gamma$ method \cite{Wolff:2003sm}.

\begin{figure}[tb]
  \centering
  \includegraphics[height=0.23\textheight]{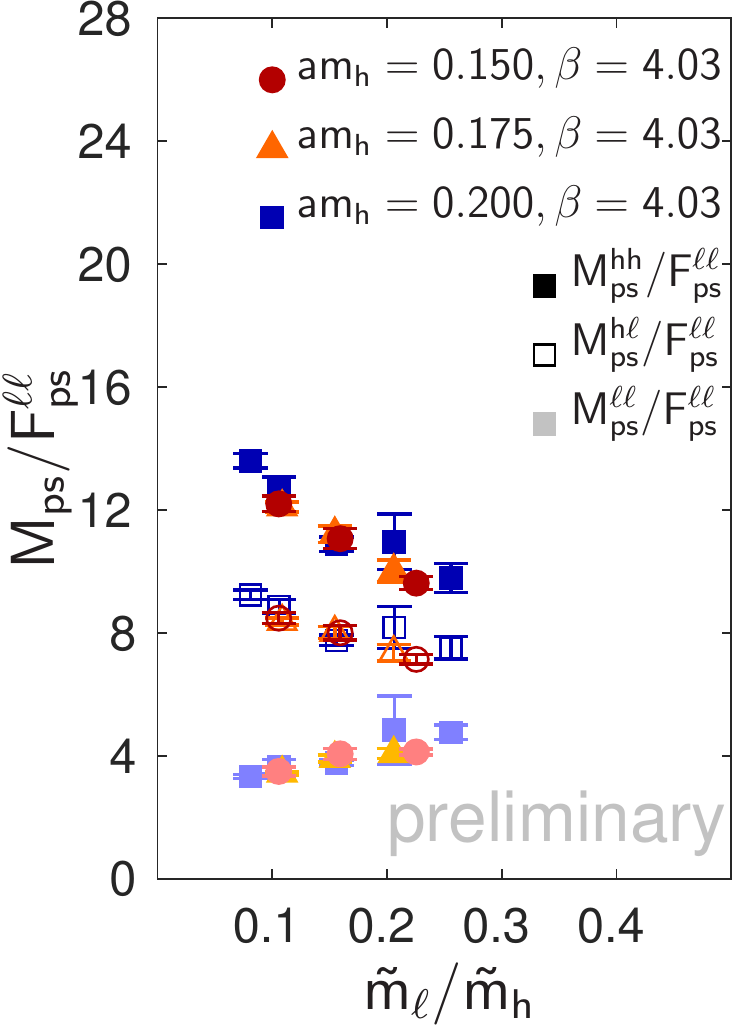}
  \includegraphics[height=0.23\textheight]{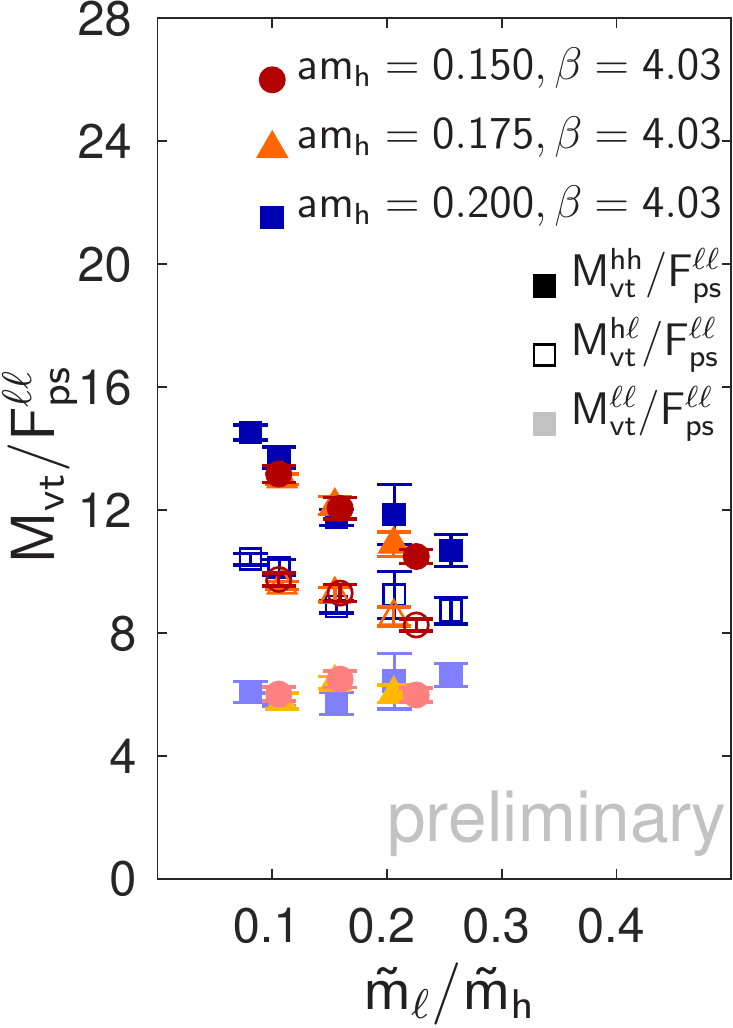}
  \includegraphics[height=0.23\textheight]{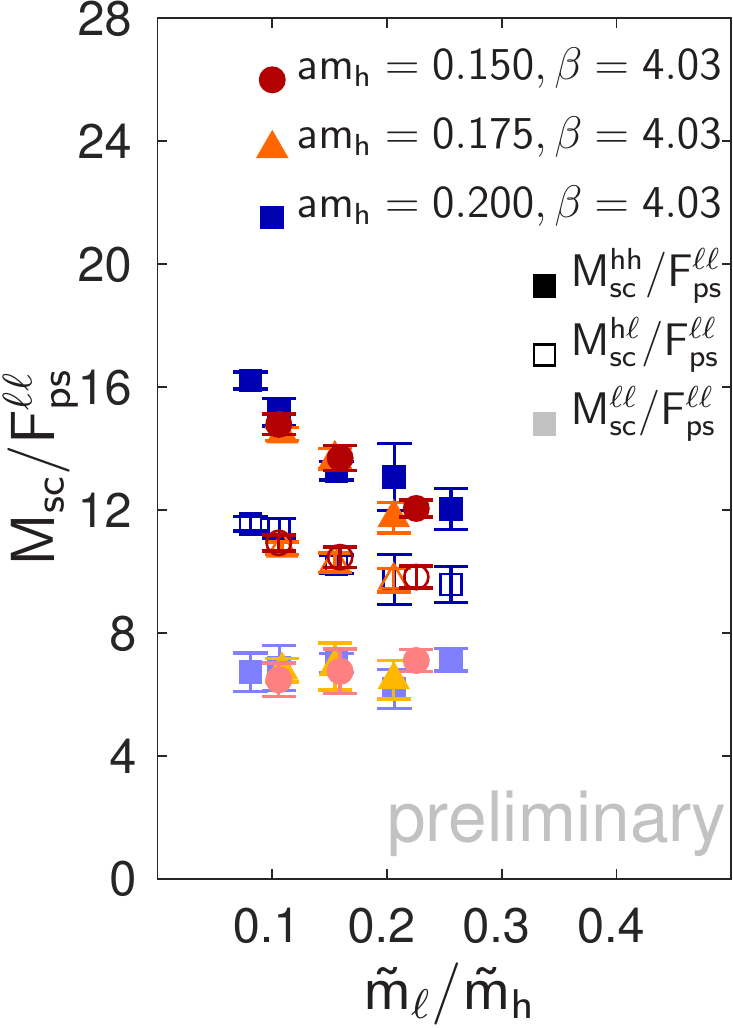}
  \includegraphics[height=0.23\textheight]{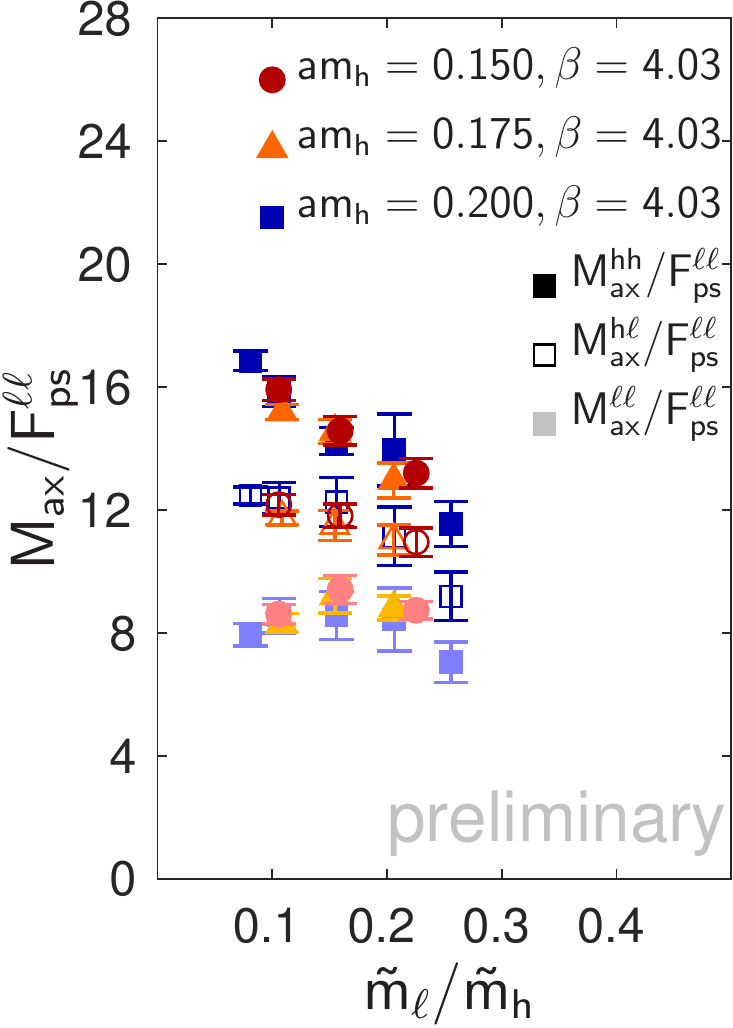}
  \caption{Low-lying connected meson spectrum obtained from light-light, heavy-light, or heavy-heavy two-point correlator functions. Shown are the pseudoscalar (ps), vector (vt), scalar (sc), and axial (ax) states in units of the pseudoscalar decay constant in the light sector $(F_\ps^{\ell\ell})$. To demonstrate hyperscaling we show the quantities as function of the ratio of light over heavy flavor mass incl.~the effect due to residual chiral symmetry breaking. Only statistical uncertainties are shown.}
  \label{fig.spectrum}
\end{figure}

We present the connected meson spectrum for pseudoscalar, vector, scalar, and axial states in Fig.~\ref{fig.spectrum}. Determining these states for light-light, heavy-light, and heavy-heavy two-point functions, we show the outcome in units of the light-light pseudoscalar decay constant $F_\ps^{\ell\ell}$. On the lattice we determine the pseudoscalar decay amplitude and obtain $F_\ps^{\ell\ell}$ after multiplying  the matching factor $Z_V$. Since for domain wall fermions $Z_V \approx Z_A$, we evaluate $Z_A$ numerically for the light-light sector and presently use that value for each ensemble to get $F_\ps^{\ell\ell}$ i.e.~$Z_A$ is obtained at finite mass and not chirally extrapolated. However, as expected for a link smeared action, $Z_A$ is one within a few percent and the mass dependence is mild. To demonstrate the hyperscaling of the ratios shown in Fig.~\ref{fig.spectrum}, we present our results as function of the light over the heavy flavor mass where the tilde indicates that the residual chiral symmetry breaking parameterized by $am_\res$ has been taken in to account e.g.~$a\tilde{m}_{\ell} = am_\ell + am_\res$. As can be seen in the plots, our three data sets with different values of $am_h$ trace out unique curves demonstrating that our results have no explicit $am_h$ dependence and thus feature conformal hyperscaling. It is noteworthy how well the very precise pseudoscalar data on the left reveal this feature.  All in all the plots in Fig.~\ref{fig.spectrum} demonstrate that the 4+6 system under investigation exhibits conformal hyperscaling. This implies that the theory with degenerate ten flavors is either conformal or extremely close to the onset of the conformal window. The running gauge coupling approaches the vicinity of a conformal FP and stays around it until the heavy flavors decouple and the light flavors break chiral symmetry spontaneously.

\begin{figure}[tb]
  \centering
  \includegraphics[height=0.165\textheight]{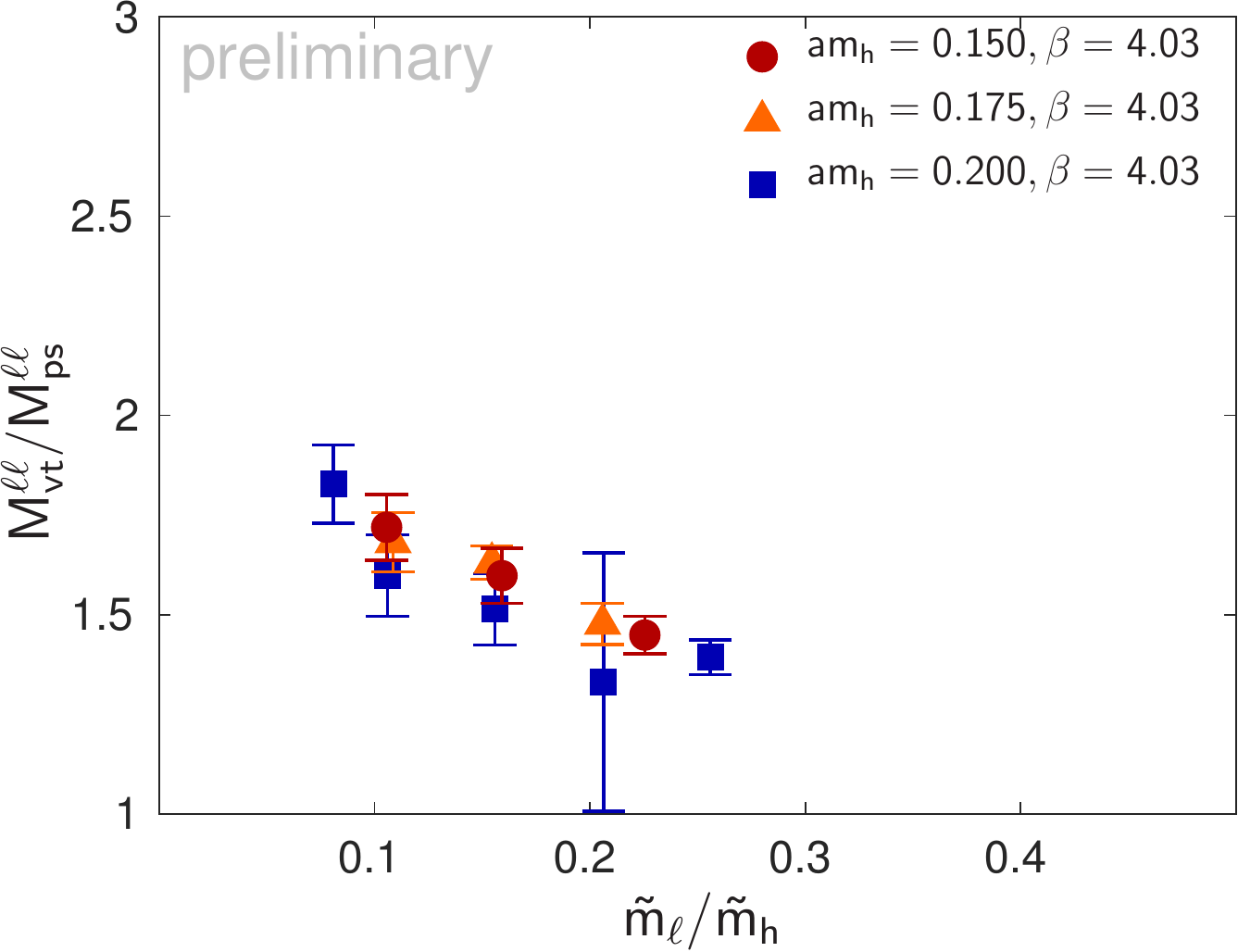} 
  \includegraphics[height=0.165\textheight]{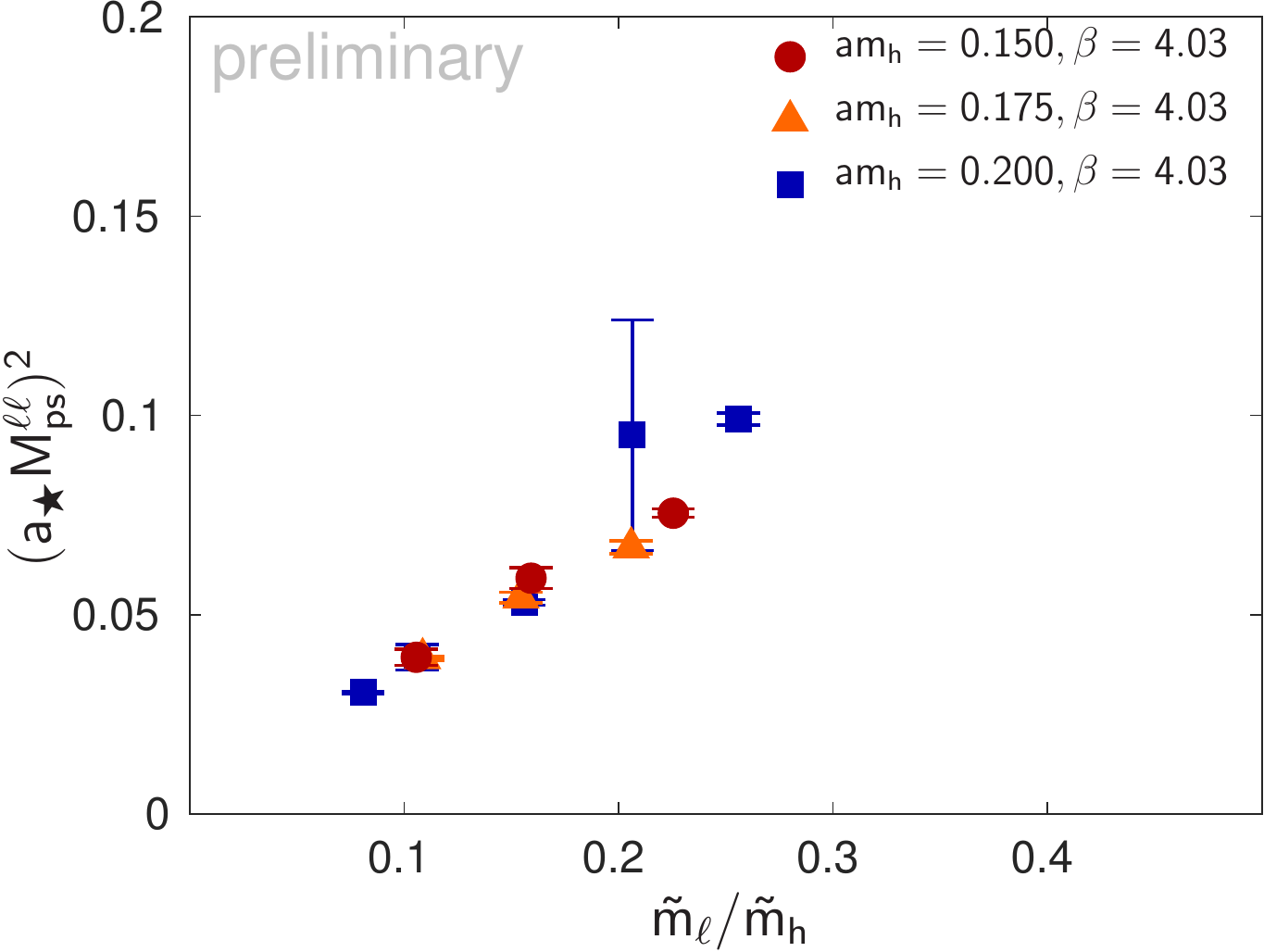}   
  \includegraphics[height=0.165\textheight]{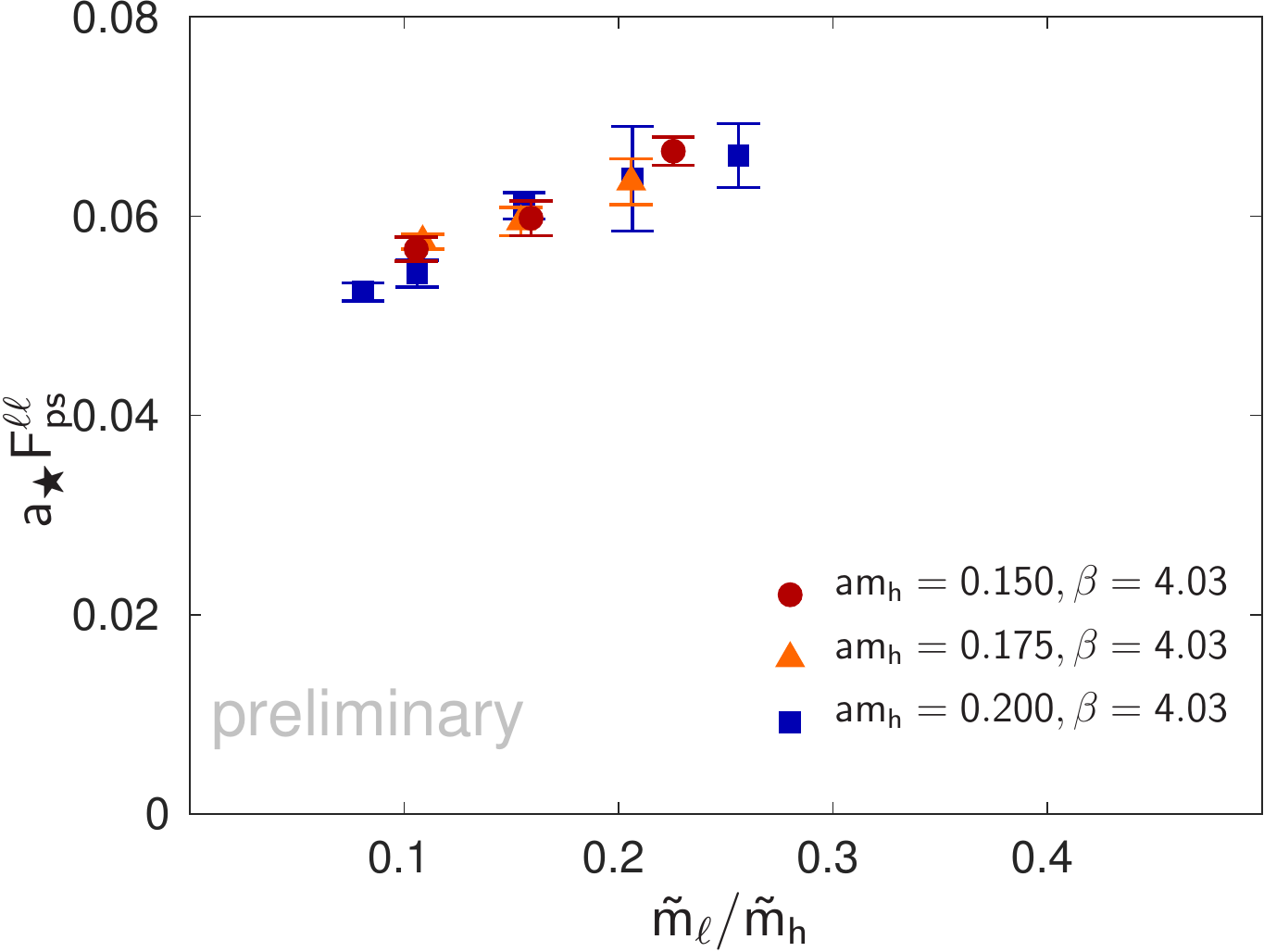}
  \caption{Chirally broken light sector. Taking the chiral limit $(am_\ell \to 0)$, we expect to observe a diverging ratio of vector the over pseudocalar mass (left panel), linear scaling of the squared pseudoscalar mass with the light flavor mass (central panel), and that the pseudoscalar decay constant approaches a finite value (right panel). Only statistical uncertainties are shown.}
  \label{fig.chiral}
\end{figure}

Next we explore the chiral properties of the light sector in Fig.~\ref{fig.chiral}. Since in a chirally broken system the mass of the pseudoscalar approaches zero in the chiral limit but the vector  meson mass remains finite, the ratio of $M_\vt^{\ell\ell}/M_\ps^{\ell\ell}$ shown in the left panel is expected to diverge in the chiral limit. Likewise the squared pseudoscalar mass is expected to be proportional to the light flavor mass (central panel) and the chiral limit of the pseudoscalar decay constant has a finite value (right panel). To account for near-conformal effects on the lattice spacing $\mathsf{a}$, we present the dimensionful quantities $M_\ps^{\ell\ell}$ and $F_\ps^{\ell\ell}$ in units of a lattice spacing $\mathsf{a}_\bigstar$. After determining the gradient flow scale $\sqrt{8 t_0}$ for each ensemble, we choose the ensemble with $m_\ell/m_h=0.015/0.200$ to be our ``reference ensemble'' ensemble and convert determinations on other ensembles to these units by multiplying $\sqrt{8t_0}\big|_{(0.015/0.200)} / \sqrt{8 t_0}\big|_{(m_\ell/m_h)}$. The large error bars on the blue data point with $m_\ell/m_h=0.040/0.200$ emphasize  finite volume effects due to using too small $16^3\time 32$ volumes.

\section{Outlook}
Our preliminary results for the connected meson spectrum clearly demonstrate that the investigated system with four light and six heavy flavors exhibits hyperscaling. In consequence we find a highly constrained hadron spectrum with only one free parameter. Since conventionally the pseudoscalar decay constant is used to set the scale, we show our results in units of $F_\ps^{\ell\ell}$. Further we demonstrate that the light sector of our model is chirally broken.

Next we intend to measure the iso-singlet scalar ($0^{++}$) in order to provide additional support that near-conformal systems exhibit a $0^{++}$ particle as lightest massive state which is much lighter than e.g.~the vector resonance. Moreover, we intend to investigate the Gell-Man-Oakes-Renner (GMOR) relation to explore whether an enhancement of the chiral condensates is observed as we drive the system closer to the conformal fixed point by lowering $am_h$. The baryon spectrum and the anomalous dimension of the baryon are straightforward to consider and will give important information for partial composite scenarios.   In addition we are generating larger $48^3\times 96$ lattices to probe deeper into the chiral limit and will add additional $24^3\times 64$ ensembles to address finite volume effects likely present in the currently used $16^3\times 32$ lattices.

\section*{Acknowledgments}\vspace{-2mm}
The authors thank their colleagues in the LSD Collaboration for fruitful and inspiring discussions. We are very grateful to Peter Boyle, Guido Cossu, Anontin Portelli, and Azusa Yamaguchi who develop the \texttt{Grid} software library providing the basis of this work and who assisted us in installing and running \texttt{Grid} on different architectures and computing centers.  A.H.~and O.W.~acknowledge support by DOE grant DE-SC0010005.

We thank the Lawrence Livermore National Laboratory (LLNL) Multiprogrammatic and Institutional Computing program for Grand Challenge allocations and time on the LLNL BlueGene/Q supercomputer. We also thank Argonne National Laboratory (ANL) for allocations through the INCITE program. Computations for this work were carried out in part on facilities of the USQCD Collaboration, which are funded by the Office of Science of the U.S.~Department of Energy and the RMACC Summit supercomputer \cite{UCsummit}, which is supported by the National Science Foundation (awards ACI-1532235 and ACI-1532236), the University of Colorado Boulder, and Colorado State University.  We thank  ANL,  BNL, Fermilab,  Jefferson Lab, LLNL, the University of Colorado Boulder, and the U.S.~DOE for providing the facilities essential for the completion of this work.

{\small
  \bibliography{../General/BSM}
  \bibliographystyle{JHEP-notitle}
}
\end{document}